\documentclass{aa}
\usepackage{epsfig}

\def\etal{{\rm et al.\thinspace}}
\def\eg{{\rm e.g.\ }}
\def\etc{{\rm etc.\ }}
\def\ie{{\rm i.e.\ }}
\def\cf{{\rm cf.\ }}

\def\spose#1{\hbox to 0pt{#1\hss}}
\def\ltsimm{\mathrel{\spose{\lower 3pt\hbox{$\sim$}}
        \raise 2.0pt\hbox{$<$}}}
\def\gtsimm{\mathrel{\spose{\lower 3pt\hbox{$\sim$}}
        \raise 2.0pt\hbox{$>$}}}
\def\Mdot{\hbox{${\dot M}$}}
\def\mdot{\hbox{${\dot m}$}}

\def\km{{\rm\thinspace km}}
\def\cm{{\rm\thinspace cm}}
\def\s{{\rm\thinspace s}}

\def\K{{\rm\thinspace K}}
\def\kmps{\hbox{${\rm\km\s^{-1}\,}$}}
\def\erg{{\rm\thinspace erg}}

\def\Msol{\hbox{${\rm\thinspace M_{\odot}}$}}

\def\pcm3{\hbox{${\cm^{-3}\,}$}}
\def\gpcm3ps{\hbox{${\rm g\pcm3\s^{-1}\,}$}}
\begin{document}
   
\title{The Evolution of Mass Loaded Supernova Remnants. II. Temperature
Dependent Mass Injection Rates}

\author{J.~M. Pittard\inst{1} \and S.~J. Arthur\inst{2} \and 
J.~E. Dyson\inst{1} \and S.~A.~E.~G. Falle\inst{3} \and 
T.~W. Hartquist\inst{1} \and M.~I.~Knight\inst{1} \and M.~Pexton\inst{3}}

\institute{Department of Physics \& Astronomy, The University of Leeds, 
        Woodhouse Lane, Leeds, LS2 9JT, UK
 \and Instituto de Astronomia, UNAM, Campus Morelia, Apartado Postal 3-72,
58090 Morelia, Michoac\/{a}n, Mexico
 \and Department of Applied Mathematics, The University of Leeds, 
        Woodhouse Lane, Leeds, LS2 9JT, UK}

\offprints{J. M. Pittard, \newline
\email{jmp@ast.leeds.ac.uk}}

\date{Received ... / Accepted ...}

\abstract{We investigate the evolution of spherically symmetric supernova
remnants in which mass loading takes place due to conductively driven 
evaporation of embedded clouds. Numerical simulations reveal significant 
differences between the evolution of conductively mass loaded and the
ablatively mass loaded remnants studied in Paper~I. A main difference is the way in which
conductive mass loading is extinguished at fairly early times, once the
interior temperature of the remnant falls below $\sim 10^{7}$~K. 
Thus, at late times remnants that ablatively mass 
load are dominated by loaded mass and thermal energy, while 
those that conductively mass load are dominated by swept-up mass and 
kinetic energy. Simple approximations to the remnant evolution,
complementary to those in Paper~I, are given.
}

\titlerunning{Evolution of Mass Loaded SNR II.}
\authorrunning{Pittard \etal}

\maketitle

\label{firstpage}

\section{Introduction}
\label{sec:intro}
The evolution and properties of supernova remnants are fundamental 
to many areas of astrophysical research. Self-similar solutions for 
various stages of the evolution have been obtained for various 
assumptions, as have unified solutions which link these together
(see Cioffi, McKee \& Bertschinger \cite{CMB1988}; Truelove \& McKee 
\cite{TM1999}). The self-similar and unified solutions are complemented 
by a wide range of numerical investigations. Amongst the problems examined
numerically are explosions in plane-stratified
media (\eg Falle \& Garlick \cite{FG1982}; Arthur \& Falle \cite{AF1991}, 
\cite{AF1993}), the 
evolution of SNRs inside pre-existing wind-driven cavities 
(Franco \etal \cite{FTBR1991}), the influence of hydrodynamic instabilities 
(\eg Chevalier \& Blondin \cite{CB1995}), and applications to the
broad emission line regions in active galactic nuclei (\eg 
Terlevich \etal \cite{TTFM1992}; Pittard \etal \cite{PDFH2001}).

In most investigations of supernova remnant evolution, the surrounding 
medium has been assumed to have a smooth density distribution. However, the
interstellar medium is known to be multi-phase (\eg McKee \& Ostriker
\cite{MO1977}), and there is clear observational evidence of engulfed 
clouds within the supernova remnant N63A (Chu \etal \cite{C1999}; 
Warren \etal \cite{WHS2002}).
As it is widely known, the injection of mass from such cool clouds into the
interior of a supernova remnant will affect its behaviour and structure.

If the clouds are assumed to be continuously distributed,
similarity solutions describing the evolution of adiabatic mass loaded 
supernova remnants can be derived for a specific functional dependence
of the mass injection. Previous work has focussed
on cases in which one mass injection process is dominant over others. 
McKee \& Ostriker (\cite{MO1977}), Chi\`{e}ze \& Lazareff (\cite{CL1981}), 
and White \& Long (\cite{WL1991}) each investigated the evolution under 
the assumption that cloud destruction is due to conductive evaporation. 
Alternatively, Dyson \& Hartquist (\cite{DH1987}) treated 
hydrodynamic ablation as the dominant process. 

A small number of 
papers based on numerical simulations of mass loaded supernova 
remnants also exist in the literature. Cowie, McKee \& Ostriker 
(\cite{CMO1981}) included the dynamics of the
clouds and found that warm clouds are swept towards the shock front and
are rapidly destroyed, while cold clouds are more evenly distributed and 
have longer lifetimes. When energy losses become important, this behaviour 
leads to the highest densities in the remnant occuring over the outer 
half radius. In contrast, when there is no mass loading, a thin dense 
shell forms at the forward shock. 
Arthur \& Henney (\cite{AH1996}) studied the effects of mass loading by
hydrodynamic ablation on supernova remnants evolving inside
cavities evacuated by the stellar winds of the progenitor
stars. They showed that the extra mass injected by embedded clumps
was capable of producing the excess soft X-ray emission seen in
some bubbles in the Large Magellanic Cloud.

The range of a supernova remnant is relevant to a number of
important areas of astronomy. It can affect the efficiency of sequential 
star formation and the global dynamics and structures of starburst 
superwinds where many remnants overlap.
The superwind in the starburst galaxy M82 must be mass loaded to account 
for the observed X-ray emission (Suchkov \etal \cite{S1996}).
Since the range and radiative energy losses of a remnant are affected by 
mass loading, it is therefore desirable to have approximations which describe 
remnant evolution and range in clumpy media.

Analogous approximations already exist for smooth media (\cf Truelove \& McKee
\cite{TM1999} and references therein). In a recent paper 
(Dyson, Arthur \& Hartquist \cite{DAH2002}, hereafter Paper~I), the first 
steps towards this goal were taken with the derivation of
range approximations for cases in which mass injection occurs 
at a constant rate, $q$, or at a rate which depends on the flow 
Mach number relative to the clump (Hartquist \etal \cite{HDPS1986}).
A mass loading rate of $qM^{4/3}$ was used for subsonic flow, where
$M$ is the Mach number of the flow. A constant mass loading rate
$q$ was used where the flow is supersonic. This prescription 
simulates mass stripping from the clump by hydrodynamic ablation. 

In this companion paper we consider mass injection from clouds into the hot 
remnant interior by conductive evaporation (\cf Cowie \& McKee \cite{CM1977};
McKee \& Cowie \cite{MC1977}). This process is potentially 
important in any astrophysical system where a hot tenuous phase coexists
with a colder denser phase (see McKee \& Cowie \cite{MC1977} for 
some examples), though it can be suppressed by magnetic fields. 
In many systems 
(including supernova remnants) the hot phase flows past the clouds, 
and they will be subject also to ablation.
Unfortunately, we currently lack even basic models of this combined 
interaction, so it is difficult to know whether one process dominates
the other, or if one process limits the other. For example, one could
imagine a scenario whereby ablation drives the initial mass loss,
increasing the effective surface area of the cloud (\eg by forming a tail),
at which point conduction takes over and leads to efficient mixing. 
In evolving objects such as supernova remnants, one process may
dominate at early evolutionary stages, while the other dominates at later 
stages. Given these potential complications and our lack of knowledge about
them, as a first approach we treat each process individually.

The outline of our paper is as follows.
In Sec.~\ref{sec:calcs} we outline our assumptions and our numerical
code; in Sec.~\ref{sec:results} we present some of our results; in 
Sec.~\ref{sec:discussion} we present simple approximations to these results;
and in Sec.~\ref{sec:conclusions} we summarize and conclude.

\section{The Calculations}
\label{sec:calcs}
As initial conditions, we take freely-expanding cold ejecta at 
constant density (typically $10^{4}$ times the ambient density, $n_{0}$),
and with a linear velocity profile. The maximum ejecta speed is 
$4000 \;\kmps$, which together with the radius of the ejecta defines 
the initial time.
We performed extensive tests with different density ratios 
to confirm that our chosen ratio is large enough not to 
influence the resulting evolution. The calculations were performed using 
an adaptive grid hydrodynamic code (see \eg Falle \& Komissarov 
\cite{FK1996,FK1998}), and we have further verified that the resolution
employed is high enough to give robust results for the global quantities
considered in this paper.

All calculations were performed in spherical symmetry, with an ejecta
mass of $10 \Msol$ and a kinetic energy of $10^{51}$~ergs (the effect
of different progenitor masses and explosion energies is commented on in 
Sec.~\ref{sec:conclusions}). The assumption
that the energy is entirely kinetic for the initial conditions is somewhat 
different to that in Paper~I, but it does not affect the resulting 
evolution (see also Gull 
\cite{G1973} and Cowie \etal \cite{CMO1981}). We modelled the intercloud 
ambient medium as a warm plasma of temperature $10^{4}$~K.
Cooling appropriate for an optically thin plasma of solar
composition and in collisional ionization equilibrium was
included. We assumed that cooling below $10^{4}$~K is balanced by
photoionization heating from the diffuse field and for that reason
impose a temperature floor of $10^{4}$~K.
For simplicity we did not include magnetic fields or shock precursors, and also
did not consider conduction between the hot remnant interior and cold 
neighbouring gas.

The rate of mass evaporation from embedded clouds is dependent on
many factors (\cf Cowie \& McKee \cite{CM1977}), including the temperature
of the hot phase, the cloud radius, and the presence of magnetic fields.
However, motions of the clouds are generally not large enough to
change the evaporation rates, and Cowie \& Songalia (\cite{CS1977}) noted that 
nonspherical clouds may be treated in an approximate way by adopting 
half the largest dimension as the radius of the cloud. 
The evaporative mass-loss rate from a single cloud 
(Cowie \& McKee \cite{CM1977}) is

\begin{equation}
\mdot =  4 \pi a^{2} n T^{1/2} (k \mu m_{\rm H})^{1/2} \phi F(\sigma_{0}), 
\end{equation}

\noindent where $a$ is the cloud radius, $n$ and $T$ are the density
and temperature of the surrounding hot phase (\ie the interior of the 
remnant), $\mu m_{\rm H}$ is the mean mass per particle, and 
$\phi$ is an efficiency factor which encapsulates the magnetic field
strength, its configuration, the cloud geometry, etc. 
($\phi=0$ corresponds to no evaporation, 
$\phi \approx 1$ to evaporation in the absence of magnetic fields -
\eg $\phi = 1.1$ for a plasma with equal electron and ion 
temperatures and cosmic abundances). $F(\sigma_{0})$ is a function 
defined in Eqs.~61 and 62 of Cowie \& McKee (\cite{CM1977}).
In the classical limit, $F(\sigma_{0}) = 2 \sigma_{0}$, where
the saturation parameter, $\sigma_{0}$, is defined as

\begin{equation}
\label{eq:sigma_0}
\sigma_{0} = \left(\frac{T}{1.54 \times 10^{7}}\right)^{2} \frac{1}{n a_{\rm pc} \phi}.
\end{equation}

\noindent Here $a_{\rm pc} = a/(1\;{\rm pc})$ is the radius of the cloud in parsecs. 
The onset of saturation occurs when $\sigma_{0}$ is of order unity, varying between
1.95 ($\phi \ll 1$) and 1.08 ($\phi = 1$) for the examples given in 
Cowie \& McKee (\cite{CM1977}). In the classical limit, $\mdot \propto T^{5/2}$
since $F(\sigma_{0}) \propto \sigma_{0}$ (\cf Fig.~4 in Cowie \& McKee \cite{CM1977}).
If saturated, $F(\sigma_{0}) \propto \sigma_{0}^{\beta}$, where 
$\beta = (1+M_{\rm s}^{2})/(6+M_{\rm s}^{2})$, and 
$M_{\rm s}$ (the Mach number of the saturated zone) is related to $\phi$ by

\begin{equation}
M_{\rm s}(1 + \frac{1}{5}M_{\rm s}^{2}) = 2\phi.
\end{equation}
 
\noindent Since $\phi$ is unlikely to exceed unity by a substantial amount, and 
has a lower limit of zero, the likely range of $\beta$ is 
$1/6 \leq \beta \leq 3/8$. Hence in the saturated limit 
$\mdot \propto T^{\alpha}$ where $5/6 \leq \alpha \leq 5/4$. 

Since the onset of saturation is dependent on the radius of the cloud
for a specified hot phase, larger clouds will tend to evaporate in the 
classical limit, while the evaporation of mass from smaller clouds will 
tend towards saturation. As the temperature and density of the remnant 
interior evolves, the radius of clouds which are just at the onset of 
saturation will 
also change. Moreover, we expect the distribution of clouds to evolve with 
time, as the timescale for the smaller clouds to completely evaporate
is shorter than the equivalent timescale for larger clouds and
in many cases the evolutionary timescale of the remnant. Further, we may
expect some differences in the physical make-up of clouds of differing
radii (\eg the distribution of cloud mass in the cold core and its 
warmer skin - \cf McKee \& Ostriker \cite{MO1977}). Finally, a full
treatment also requires specifying the number of clouds
per unit volume, which is likely to
be dependent on the local environment (\eg the cloud
spectrum inside a starburst superwind is likely to be somewhat different
to that in our local ISM). McKee \& Ostriker (\cite{MO1977}) argued that
mass loading in the Galactic ISM would be dominated by the smaller clouds.

Addressing all of these issues is beyond the scope of the 
present paper, so we choose instead the following simplification. 
We assume that we can specify a temperature for the onset
of saturation, $T_{\rm sat}$, which is an average over both the 
cloud distribution and time. We adopt $T_{\rm sat} = 10^{7}$~K for all of 
our calculations. Given the uncertainties introduced by 
this assumption (and those in the original work of Cowie \& McKee
\cite{CM1977}) it seems unrealistic to specify a complicated dependence
for the mass evaporation rate per unit volume above $T = T_{\rm sat}$,
and we therefore assume that it is independent of $T$.
Tests with other values of $T_{\rm sat}$ have shown that our results 
only become sensitive to the value of $T_{\rm sat}$
when $T_{\rm sat} < 10^{7} \sqrt{n_{0}} \K$ ($n_{0}$ ($\pcm3$) 
is the intercloud ambient number density). Hence our results with 
$n_{0} \leq 1 \pcm3$ are robust (since a lower value of $T_{\rm sat}$ 
implies that most of the mass loading is from a population of tiny clouds 
($a_{\rm pc} < 0.42/\phi$) whereas the smallest cloud radii in models
of the ISM by McKee \& Ostriker (\cite{MO1977}) are $\sim 0.38$~pc).
On the other hand, for $n_{0} > 1 \pcm3$, our results are dependent 
on our chosen value of $T_{\rm sat}$. However, as $T_{\rm sat}$ will
depend on details such as clump lifetimes and number distributions,
which are problem specific parameters, we simply note that our results
begin to suffer from loss of generality in this regime.
 
Our final assumptions are: i) that the intercloud 
spacing is small compared to the scale over 
which the properties of the remnant varies (so that a continuous mass source
term can be used), ii) that injection takes place
at zero velocity relative to the global flow, and iii) 
that the injected gas has zero internal energy. The full set of
gas dynamic equations is then

\begin{equation}
\frac{\partial \rho}{\partial t} + \frac{1}{r^{2}}\frac{\partial}{\partial r} 
r^{2}\rho u = q,
\end{equation}

\begin{equation}
\frac{\partial \rho u}{\partial t} + \frac{1}{r^{2}}\frac{\partial}{\partial r} 
r^{2}(p + \rho u^{2}) = \frac{2p}{r},
\end{equation}

\begin{equation}
\frac{\partial e}{\partial t} + \frac{1}{r^{2}}\frac{\partial}{\partial r} 
\left[r^{2} u (e + p)\right] = -L,
\end{equation}

\noindent where the symbols $\rho$, $u$, and $p$ have their usual meanings,
$e = p/(\gamma-1)+0.5\rho u^{2}$ is the total energy, and $L$ is the
radiative cooling term. $q$ is the mass loading rate per unit volume 
and time. 

As in Paper~I, we define a fiducial mass loading rate

\begin{equation}
\label{eq:q0}
q_{0} = \frac{6 \rho_{0}}{5 t_{\rm sf}},
\end{equation}

\noindent which gives the mass flux through the blast wave divided by the
remnant volume at the onset of thin shell formation for
Sedov expansion. Here, $\rho_0$ is the
intercloud ambient density and $t_{\rm sf}$
is the timescale for thin shell formation for a remnant propagating into a 
uniform ambient medium given by Franco \etal 
(\cite{FMATT1994}),

\begin{equation}
\label{eq:tsf}
t_{\rm sf} = 2.87 \times 10^{4} E_{51}^{3/14} n_{0}^{-4/7} \;\;{\rm yrs},
\end{equation}

\noindent where $10^{51} E_{51}$ is the explosion energy. 
In these units, $q_{0} = 1.32 \times 10^{-36} E_{51}^{-3/14} n_{0}^{11/7}$ (with
mean mass per particle, $\mu =0.6$).

Following the above discussion, we define the mass loading rate 
due to conduction as

\begin{eqnarray}
\label{eq:q}
q = & 316f'q_{0}         & (T > 10^{7}~{\rm K}), \nonumber \\
q = & f'q_{0} (T/T_{6})^{5/2} \hspace{10mm} & (10^{5}~{\rm K} < T < 10^{7}~{\rm K}), \nonumber \\
q = & 0              & (T < 10^{5}~{\rm K}),
\end{eqnarray}
 
\noindent where $T_{6} = 10^{6}$~K (\eg Cowie \etal \cite{CMO1981}). 
Thin shell formation occurs when the temperature immediately
behind the leading shock is about $10^{6}$~K. At low temperatures, 
mass loading is so weak that the precise temperature cutoff 
is unimportant. Defining $f'$ as we have in Eq.~\ref{eq:q} ensures that   
mass loading starts to influence the dynamics and
evolution of the remnant if $f' \sim 1$. To be more specific, 
low values of $f'$ mean that mass loading will not
dominate in the remnant before the time of onset of thin shell
formation, while $f' > 1$ indicates that the evaporated mass
becomes important in the remnant before $t_{\rm sf}$ is reached.
Readers who refer to Paper~I will note that in that paper we 
defined a parameter $f$ which is somewhat analogous to $f'$. In an
ablatively mass loaded remnant the ablated mass is comparable to the 
swept-up mass at time $t_{\rm sf}$ if $f \sim 1$. However, under the
assumptions we have made, in very young remnants in which the temperature
is everywhere above $10^{7}$~K, $f \sim 316 f'$ for the mass loading
rate in an ablatively loaded remnant to be comparable to that in a 
conductively loaded remnant for which $E_{0}$ and $n_{0}$ are the same.

$f'$ is physically related to the number density of clouds and the rate
at which mass is evaporated from these. However, from Eq.~13 of
Cowie \etal (\cite{CMO1981}), and since
$q_{0} \propto E_{51}^{-3/14} n_{0}^{11/7}$, we find that

\begin{equation}
\label{eq:fdash_propto}
f' \propto \phi a_{\rm pc} N_{\rm c} E_{51}^{3/14} n_{0}^{-11/7},
\end{equation}

\noindent where $N_{\rm c}$ is the number density of clouds.
Thus we might expect real remnants to be better described by small $f'$ 
when they evolve in environments with high ambient densities, and larger $f'$ 
when they evolve in environments with low ambient densities. 
We note that if the distribution of clouds evolve with time, $f'$ itself 
may evolve with time, though such an effect is not investigated in
the models presented in this paper.

We note here, that differences exist between
this work and Paper~I. In the calculations presented here, mass 
loading is permitted throughout the remnant, including the ejecta material, whereas 
in Paper~I it is switched off in the ejecta.
This causes differences when most of the mass loading 
occurs before the remnant has swept up much intercloud mass (\ie when
$f'$ and $n_{0}$ are large). 
A further difference is the assumed temperature
of the ambient gas. In Paper~I, a temperature of 100~K was used, whereas here we use 
$10^{4}$~K. This will have negligible consequences until near the merger
of the remnant with the surrounding gas. We note that in the specific case of 
superwind generation, merger may take place with gas at even higher temperatures.

We have computed models with $f'=0,0.316,3.16,31.6,316$, and
ambient densities $n_{0}=0.01,0.1,1,10,100\;\pcm3$. 
Although there is an indication that high values of $f'$ will 
be favoured when remnants evolve in surroundings with low ambient densities,
$f'$ also scales with the number density of clouds, 
and thus could be completely independent of the density of the 
intercloud medium. Hence we explored ($f',n_{0}$) parameter space.

The remnants are evolved until the blast front degenerates into an acoustic wave.
We adopted the same definitions for the ends of the free expansion and
Quasi-Sedov-Taylor phases as given in Paper~I. These are respectively when the 
internal energy reaches 60\% of the initial energy (FE) and when 
50\% of the initial energy has been radiated (QST). For $f' \ltsimm 3.16$, the mass loading 
dominates in the remnant at times before the onset of thin shell formation, 
$t_{\rm sf}$, though is always decreasing in importance at this point
(\ie the ratio of loaded mass to swept-up mass is falling). For $f' > 3.16$ 
the mass loading also remains dominant at the time of the end of the 
QST stage, $t_{\rm QST}$. Finally, as in Paper~I, we note the caveat that we are 
considering a purely general case and that, in practice, $f'$ is determined by 
the actual physical situation being studied. We therefore ignore details such
as clump lifetimes and spatial distributions, which only introduce additional,
problem specific, parameters.

\section{Results}
\label{sec:results}
We note that all of the results presented in this paper and Paper~I
are for a particular set of explosion parameters, namely the
canonical values $E = 10^{51}\;\erg$
and $M = 10 \Msol$. Where results for hydrodynamic ablation or
constant mass loading are presented these have been calculated with
the same code as used in the conductive mass loading cases but for the
mass loading prescription given in Paper~I. 

\subsection{Remnant Expansion Speeds and Ranges}
\label{subsec:exprange}
In Fig.~\ref{fig:fig1} we show the remnant expansion speed and radius as
functions of time for calculations with zero mass loading ($f'=0$), substantial
mass loading ($f' = 3.16$), and high mass loading ($f'=31.6$), and for 
ambient densities of 
$n_{0} = 0.01\;\pcm3$ and $n_{0} = 100\;\pcm3$. Mass loading causes 
a reduction in expansion velocity and range at intermediate 
ages, but there is convergence in speeds and ranges at later times, 
regardless of the value of $f'$ or the ambient density.
This is in contrast to remnants which mass load through ablation (Paper~I)
where the reductions in expansion speed and remnant range persist
right through to the point of merger with the ambient medium.

This dynamical behaviour can be understood as follows.  At early
times, the highest temperatures are in the shocked ISM material so
mass loading is highest there. By conservation of momentum, the
addition of mass to this gas reduces its velocity and also has the
effect of increasing the pressure (by conservation of energy). This
causes the shocked material to brake harder and leads to an overall
reduction in the expansion velocity of the remnant. Once the ejecta
have completely thermalized, the highest temperatures are in the
shocked ejecta material and most of the mass is added here, increasing
the density and pressure in this gas. At this stage the rate of
decrease of the expansion velocity is less in the conductively mass
loaded remnants than in the zero mass loading case because of the
higher pressure in the remnant interior. As the blast wave
decelerates, the postshock temperatures and hence the conductive mass
loading rates decrease, so during this stage the swept-up mass becomes
more important. Eventually, the swept-up mass dominates the dynamics
in the postshock region and the remnant evolution tends towards the
zero mass loading case.

\subsection{Mass Loading History}
\label{subsec:mlhist}
Fig.~\ref{fig:fig2} shows the ratio of injected to swept-up mass as a
function of time for various values of $f'$ and ambient densities
$n_{0}$.  The top panels show the evolution when mass is injected
through hydrodynamic ablation, while the bottom panels show the
results obtained when mass injection occurs by conductive evaporation.
It is immediately obvious from this figure that these mass loading
prescriptions yield vastly different behaviour.

When mass loading occurs through ablation, the injected mass becomes
increasingly dominant over the swept up mass as time progresses.  The
limit of the ablative mass loading is a constant mass loading rate $q$
(corresponding to supersonic flow throughout the remnant), so that the
injected mass increases as $\Mdot_{\rm load} \approx 4\pi R(t)^{3}
q/3$, whereas the rate of increase of the swept-up mass is $\Mdot_{\rm
  swept} = 4 \pi R(t)^{2} \rho_{0} dR(t)/dt$, so that
\begin{equation}
\frac{\Mdot_{\rm load}}{\Mdot_{\rm swept}} \propto \frac{R(t) q}
{{\dot R}(t) \rho_{0}} \propto f't/t_{\rm sf},
\end{equation}
where we have used Eq.~\ref{eq:q0} and approximated the remnant
expansion speed as $\dot{R}(t) \propto R(t)/t$. Hence, as the ratio of
the {\em rates} increases approximately linearly with time, we expect
that the ratio of the injected to swept-up {\em mass} will also
increase approximately linearly with time. This is indeed the
behaviour that we see in Fig.~\ref{fig:fig2} (a,b). Furthermore, for a
given $f'$, we expect that the ratio of the loaded mass ($M_{\rm
  load}$) to the swept-up mass ($M_{\rm swept}$) as a function of time
(in units of $t_{\rm sf}$) will be independent of $n_{0}$, as is
indeed seen (Fig.~\ref{fig:fig2} (b)). 
With reference to Eq.~\ref{eq:fdash_propto}, however,
``mathematical'' independence of $n_{0}$ (at constant $f'$) does not 
imply ``physical'' independence (\ie keeping constant the other
physically meaningful parameters, like $\phi$, $a_{\rm pc}$, 
$N_{\rm c}$, $E_{51}$).

When mass loading occurs via conductively-driven evaporation, on the
other hand, the behaviour of $M_{\rm load}/M_{\rm swept}$ as a
function of $t$ is quite different (\cf Fig.~\ref{fig:fig2} a,c).
Although the injected mass exceeds the swept-up mass at intermediate
times, as the remnant ages the rate of mass loading becomes
negligible, and the swept-up mass gradually dominates. In
Fig.~\ref{fig:fig2} (c,d), the swept-up mass is, in all cases, greater
than the injected mass at the point when the remnant merges with the
ambient medium. 

\begin{figure*}
\begin{center}
\psfig{figure=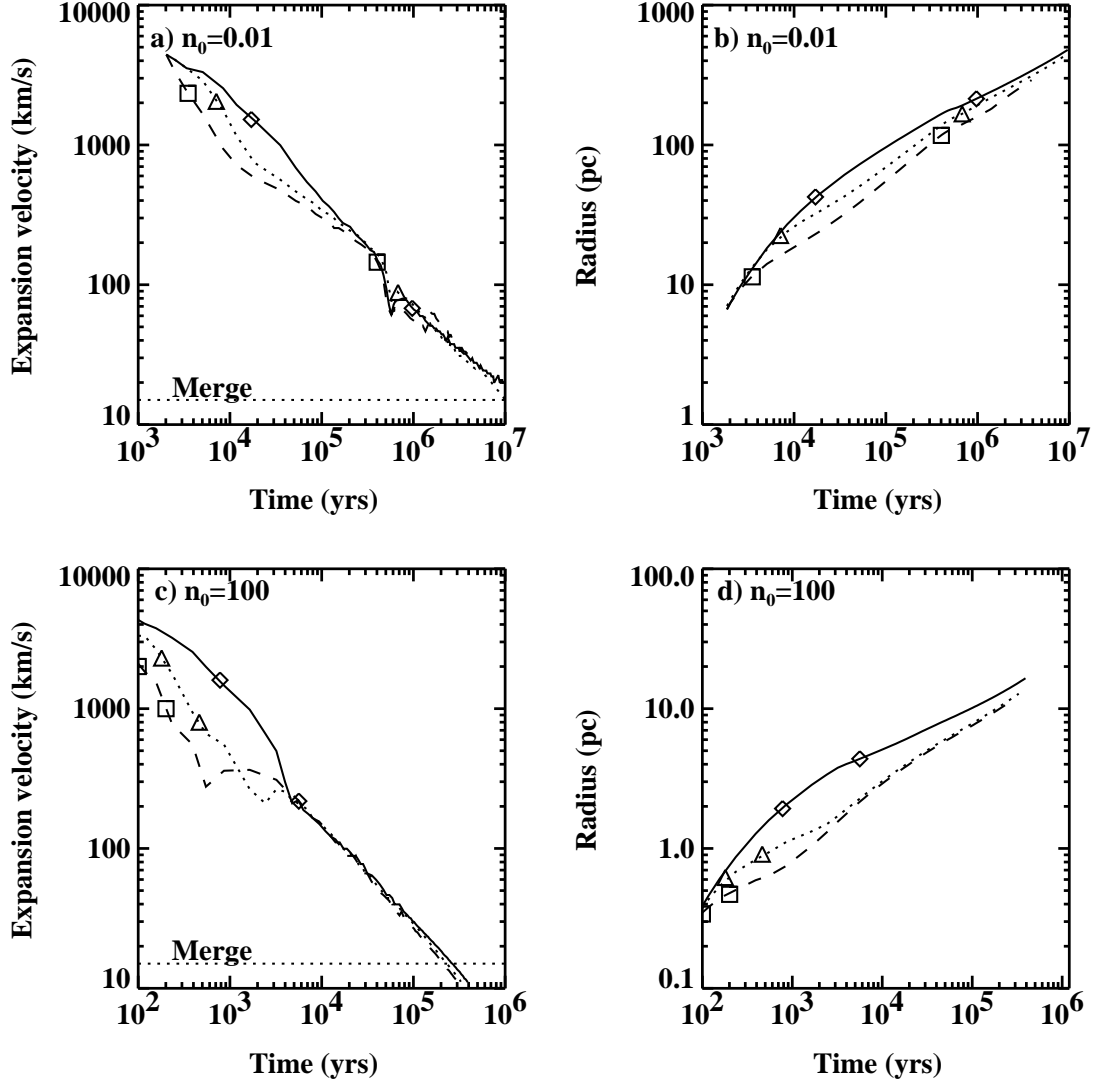,width=15.0cm}
\end{center}
\caption
[]{Expansion velocity ($\kmps$) and radius (pc) versus time (yrs)
for remnants ($E = 10^{51}\;\erg$, $M = 10 \Msol$) 
expanding into an ambient density of $n_{0} = 0.01 \pcm3$ 
and $n_{0} = 100 \pcm3$. The temperature of the ambient medium is 
$10^{4}\;{\rm K}$, and its sound speed is $15 \kmps$. The solid line 
has $f'=0$ (no mass loading), while the dotted and dashed lines have 
$f'=3.16$ (substantial mass loading) and $f'=31.6$ (high mass loading) 
respectively. The diamonds, triangles and squares indicate the end of
the FE and QST stages for $f'=0,\,3.16,\,31.6$ respectively.}
\label{fig:fig1}
\end{figure*}

\begin{figure*}
\begin{center}
\psfig{figure=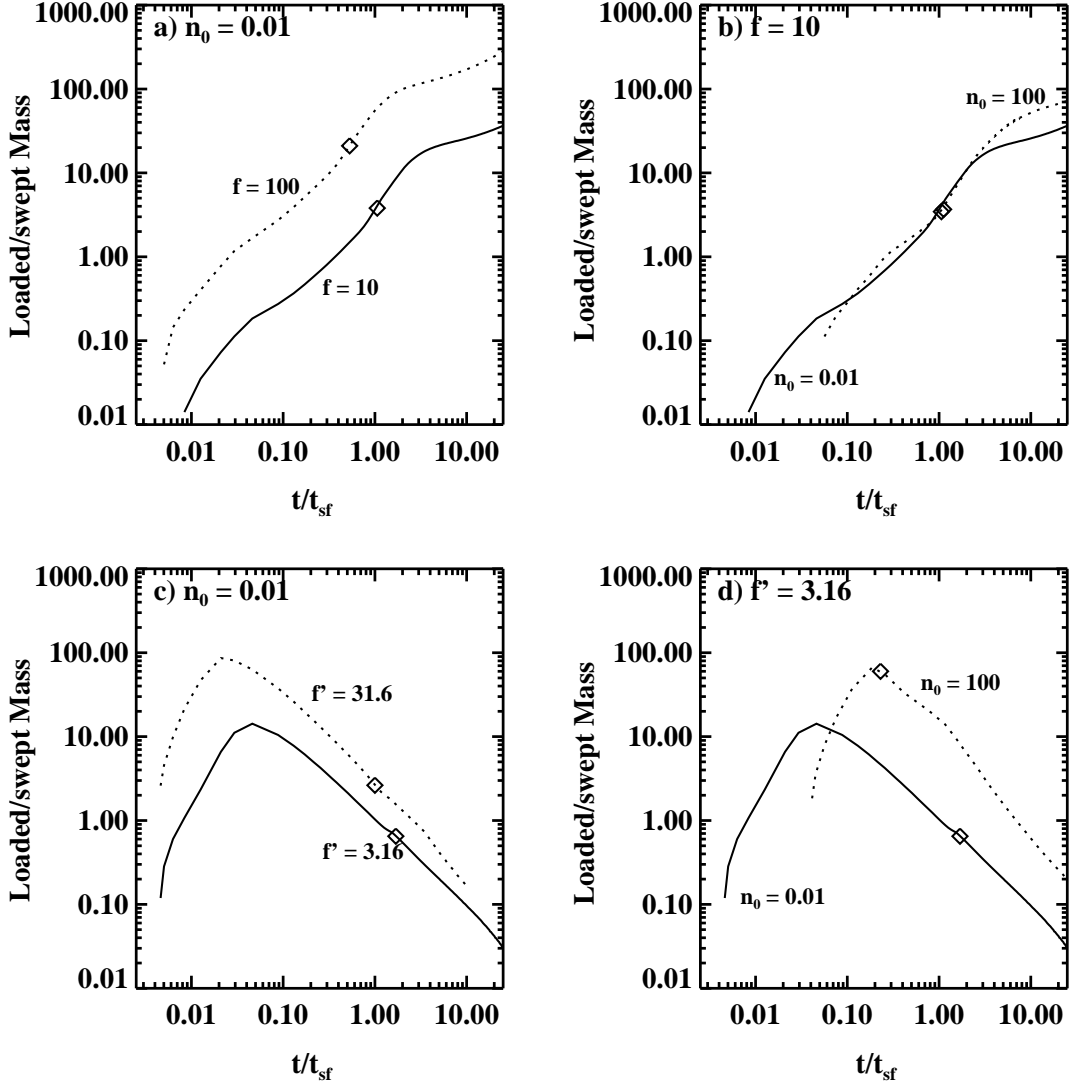,width=15.0cm}
\end{center}
\caption[]{The ratio of evaporated or ablated mass to swept-up mass as a 
function of time
(normalized to the time of shell formation \cf Eq.~\ref{eq:tsf}) 
for mass loading by hydrodynamic ablation (top panels) or by conductive
evaporation (bottom panels). $t_{\rm sf} = 4 \times 10^{5}$~yr ($n_{0}=0.01$)
and $2 \times 10^{3}$~yr ($n_{0}=100$). The diamonds mark the end of the 
QST stage, and almost overlap in b).}
\label{fig:fig2}
\end{figure*}

This qualitative difference in the time-dependent behaviour of the
mass loading between the conductively driven evaporation and ablation
cases accounts for the differences in the evolutionary behaviour
displayed in Fig.~\ref{fig:fig1} of this paper and the evolution of
the remnants discussed in Paper~I.  For the conductive case, the
mass loading is extinguished at relatively early times and hence does
not have much effect over the later stages of the remnant
evolution. Once mass loading is ``switched off'' the ratio 
$M_{\rm load}/M_{\rm swept}$ depends solely on $M_{\rm swept}$ and
thus is {\em not} independent of $n_{0}$.

\subsection{Mass and Energy Fractions}
\label{subsec:mefrac}
Fig.~\ref{fig:fig3} shows the time variation of the mass and energy
fractions for $n_{0} = 0.01\;\pcm3$ and $f' = 0$, 3.16 and 31.6.
Although our initial conditions specify the remnant energy as being
entirely kinetic at early times, a significant fraction is quickly
thermalized.

Values of $f' = 3.16$ and 31.6 correspond to $f$ values of $10^4$ and
$10^5$, respectively. These are much higher than the $f$ values
discussed in Paper~I. The reason for this choice is to approximately
match the loaded to swept-up mass ratio in the two different mass
loading scenarios at the onset of thin-shell formation,
$t_\mathrm{sf}$, i.e., $M_\mathrm{load}/M_\mathrm{swept} \approx 4$ in
both cases. Since conductive evaporation is saturated only at early
times in the remnant evolution (\cf Fig.~\ref{fig:fig3}c and e) and
becomes negligible once average temperatures fall below $10^6$\,K,
while hydrodynamic ablation becomes more important as the remnant's
volume increases, obviously far higher values of $f$ are necessary in
the conductive evaporation case to give
$M_\mathrm{load}/M_\mathrm{swept} \approx 4$ at $t_\mathrm{sf}$.

With $f'=0$ (no mass loading) the thermal energy fraction peaks at
approximately 0.72 \ie the value for a Sedov-Taylor remnant expanding
into a uniform medium (\cf Fig.~\ref{fig:fig3}b). Once the age of the
remnant is comparable to $t_{\rm sf}$, the remnant begins to radiate
its thermal energy, and the thermal energy fraction falls as the
kinetic energy fraction rises. Mass injection speeds up these
processes, and the thermal energy fraction first overshoots (see
\S~\ref{subsec:exprange}), then undershoots the Sedov-Taylor value
(Figs~\ref{fig:fig3}d,f). The decrease in the thermal energy fraction
occurs not long after the end of the FE stage. This is long before
radiative losses become significant, and the remnants are at this
stage cooling by adiabatic expansion, such that thermal energy is
converted into kinetic energy. This is despite the countering effect
of mass loading, which at any given time tends to increase the thermal
energy fraction relative to the kinetic energy fraction. However, we
note that the thermal energy fraction begins to decline when the mass
fraction of evaporated material starts to drop off and thus is simply
a manifestation of the fact that mass loading is ceasing to be an important
process in these remnants long before the onset of radiative
cooling. In contrast, for the ablation cases discussed in Paper~I, it
was found that since mass loading becomes more important for the
remnant as time goes on, the thermal energy fraction
continues to increase right up until the end of the QST stage.

At very late times the energy in a SNR with mass loading due to
conductive evaporation is predominantly kinetic, whereas in Paper~I it
was found that in the final stages of a SNR with mass loading due to
ablation the continued mass loading ensures that thermal energy
dominates.

\subsection{Evolution of Physical Quantities}
In Fig.~\ref{fig:fig4} we show the density, pressure, temperature, and
velocity distributions at specific times just before and after the
predicted onset of thin-shell formation for models with $n_{0}=0.01$
and $f'=0$ and $f'=3.16$.  The ``structure'' at small radii (where the
ejected mass is) seen in the density and temperature plots is a
consequence of the initial conditions and imposed spherical symmetry
(which requires a reflection condition at $r = 0$) but does not
significantly affect global properties (see \eg Cioffi \etal
\cite{CMB1988}).

For $f'=0$ the distributions of density \etc in the external part of
the remnant (where the swept up mass is) correspond to
the standard Sedov-Taylor solution (although, as in Cioffi \etal
\cite{CMB1988}, at high ambient densities the remnant becomes
radiative before fully entering this phase).  The thin-shell formation
time, $t_{\rm sf}$, for this ambient density is $4 \times 10^{5}$~yr
(from Eq.~\ref{eq:tsf}). Inspection of Fig.~\ref{fig:fig4} (in
particular, the pressure, which drops markedly in the region where strong
radiative cooling starts) shows that our simulations are consistent
with this value.

The addition of mass through the conductively driven evaporation of
embedded clouds significantly alters the properties of the remnant,
most obviously causing the interior density of the remnant to increase
with respect to the no mass loading case. At early times, mass loading
occurs mainly in the centre of the remnant (in the shocked ejecta
region). Once temperatures here fall below $10^6$\,K (due to adiabatic
expansion), mass loading occurs mainly in the region of hot post-shock
gas behind the blast wave, creating a ``thick-shell'' morphology (\cf
Fig.~\ref{fig:fig4}b), resembling that of Cowie \etal (\cite{CMO1981})
and Dyson \& Hartquist (\cite{DH1987}).

\subsection{Volume-Averaged Mass Loading Properties}
\label{avprop}
In Fig.~\ref{fig:fig5}~(a) we show the remnant-averaged mass-injection
rate (rate of mass evaporation divided by the volume of the remnant)
as a function of time.  At early times, the entire remnant is above $T
= 10^7$\,K, hence conductive evaporation is saturated and the
remnant-averaged mass loading rate is approximately constant as the 
remnant expands (the small rise is caused by the increasing fractional
volume of the hot shocked gas in the remnant). Once
the average remnant temperature drops below $10^{7}\;{\rm K}$ the mass
evaporation rate is no longer saturated and $\dot{\rho}$, the rate of
mass evaporation per unit volume, varies with time as $\dot{\rho}
\propto t^{-2}$. This matches the behaviour found by Chi\`{e}ze \&
Lazareff (\cite{CL1981}), who adopt $\dot{\rho} \propto T^{5/2}$, for
heavily mass loaded remnants.  This is a steeper dependence than the
$\dot{\rho} \propto t^{-1}$ dependence found by White \& Long
(\cite{WL1991}), which is a consequence of their particular
description for mass loading ($\dot{\rho} \propto T^{5/6}$).

From Fig.~\ref{fig:fig5}b and Fig.~\ref{fig:fig5.5} we see that the
majority of the mass evaporation from the embedded clouds occurs after
the average temperature of the remnant drops below $10^{7}\;{\rm K}$.
Once the average remnant temperature drops below $10^5$\,K mass
loading is effectively ``switched off'', the remnant averaged mass
injection rate drops sharply, leaving the power-law dependence on
time, and the quantity of evaporated mass remains constant in the
remnant from this time onwards. Interestingly, this quantity does not
depend linearly on the mass loading factor $f'$ (Fig.~\ref{fig:fig5}b).
Roughly half the evaporated mass is loaded after the average
temperature has dropped below $10^{6}\;{\rm K}$ for the model with
$n_{0}=0.01\;\pcm3$ and $f'=3.16$. Hence our results should not be
strongly affected by the precise value of the temperature $T_{\rm
  sat}$ at which conductive evaporation becomes saturated, or by the
initial conditions that we specify.

\subsection{End of the QST Stage}
In Fig.~\ref{fig:fig6}, the ratio of the time at which the total
energy of the remnant falls to one half of its initial energy,
$t_{1/2}(f',n_{0})$, to the shell formation timescale, $t_{\rm sf}
(n_{0})$, defined in Eq.~\ref{eq:tsf}, is shown as a function of the
mass loading parameter $f'$ for the five different values of the
ambient density used. This figure shows far more variation with
$n_{0}$ than the equivalent figure in Paper~I, where the curve for the
different values of the mass loading parameter lie on top of each
other. In the conductively driven mass loading case the mass loading
rate is sensitive to the thermal structure throughout the remnant, and
radiative cooling, when it begins, occurs in a larger volume of gas
(because of the ``thick-shell'' morphology). The radiative cooling
rate thus depends on the density and temperature of the gas in this
extended region, which have a non-simple relation to the mass loading
factor $f'$. In contrast, in the ablation cases discussed in Paper~I,
mass loading occurs in the narrow region just behind the blast wave,
and so the amount of radiative cooling in this zone (which ultimately
defines $t_\mathrm{QST}$) depends directly on the factor $f$.

\begin{table}
\begin{center}
\caption{The total mass, $M_{\rm FE} (\Msol)$, in the remnant at the 
end of the 
free expansion stage for conductive mass loading. Apart from models with
$f'=0.316$, the mass is relatively invariant with $n_{0}$, being 
$\approx 56\;\Msol$ for $f'=0$ and $\approx 37\;\Msol$ for $f' \geq 3.16$.
$E = 10^{51}\;\erg$, $M = 10 \Msol$.}
\label{tab:mass_fe}
\begin{tabular}{rrrrrr}
\hline
 & \multicolumn{5}{c}{$f'$} \\
$n_{0}$ & 0.0 & 0.316 & 3.16 & 31.6 & 316 \\
\hline
0.01 & 56.4 & 45.6 & 38.1 & 36.2 & 36.4 \\
0.10 & 56.4 & 43.3 & 37.4 & 36.9 & 37.6 \\
1.00 & 56.0 & 42.4 & 37.6 & 36.2 & 36.2 \\
10.0 & 55.6 & 41.6 & 36.9 & 37.6 & 37.6 \\
100.00 & 56.0 & 39.3 & 37.6 & 37.0 & -- \\
\hline
\end{tabular}
\end{center}
\end{table}

\begin{table}
\begin{center}
\caption{The radius (pc) of a conductively mass loaded remnant at the 
end of the Quasi-Sedov-Taylor stage. $E = 10^{51}\;\erg$, $M = 10 \Msol$.}
\label{tab:rad_qst}
\begin{tabular}{rrrrrr}
\hline
 & \multicolumn{5}{c}{$f'$} \\
$n_{0}$ & 0.0 & 0.316 & 3.16 & 31.6 & 316 \\
\hline
0.01 & 214 & 201 & 167 & 117 & 50.7 \\
0.10 & 85.8 & 72.6 & 56.4 & 23.9 & 7.97 \\
1.00 & 32.8 & 21.8 & 11.7 & 4.45 & 2.00 \\
10.0 & 12.4 & 5.67 & 2.77 & 1.32 & 0.67 \\
100.00 & 4.35 & 1.74 & 0.89 & 0.45 & 0.23 \\
\hline
\end{tabular}
\end{center}
\end{table}

\begin{figure*}
\begin{center}
\psfig{figure=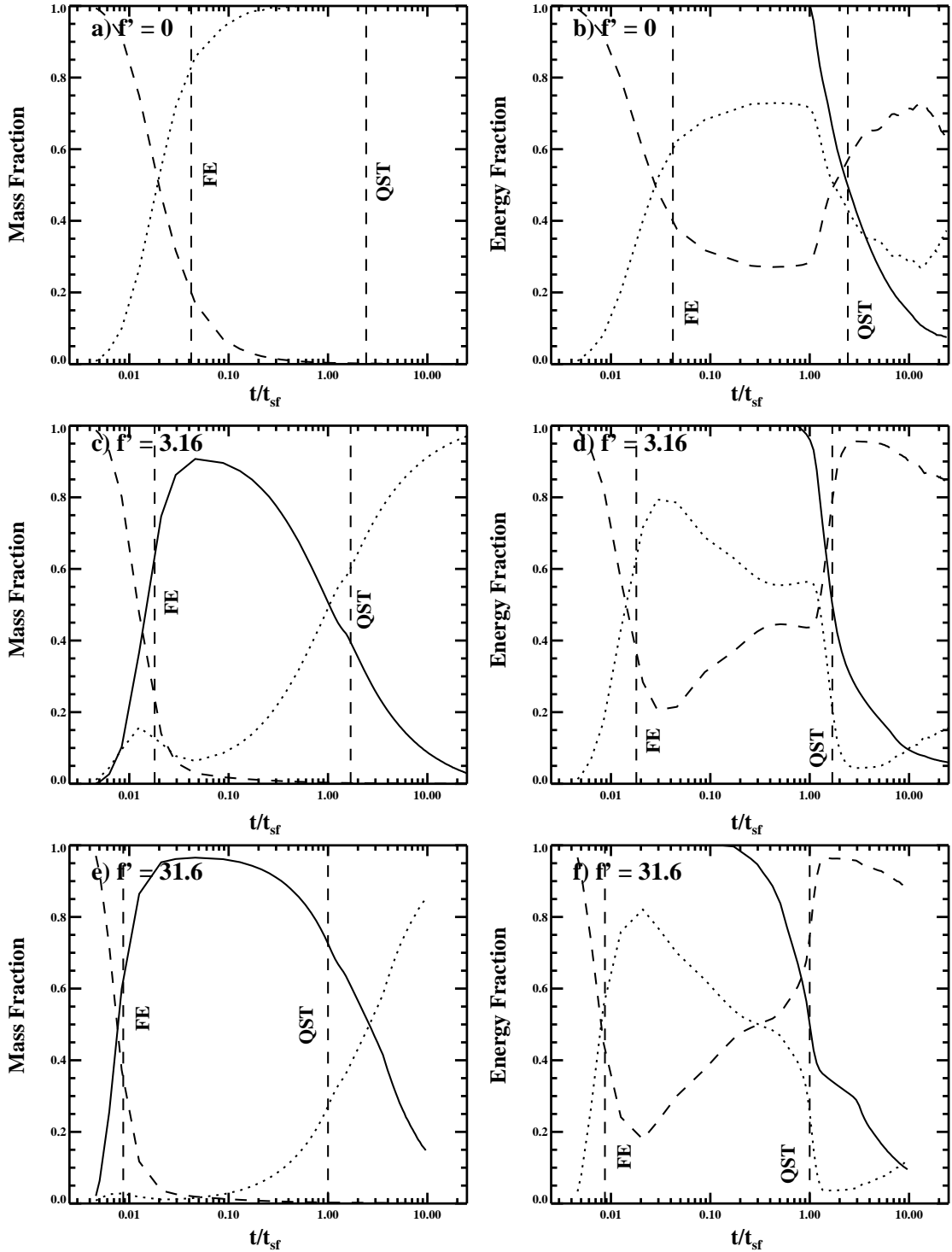,width=15.0cm}
\end{center}
\caption[]{Mass fractions (left panels) and energy fractions (right panels) 
as functions of the remnant age for $f' = 0, 3.16, 31.6$ ($f = 0,10^{4},10^{5}$)
and $n_{0} = 0.01$ ($t_{\rm sf} = 4 \times 10^{5}$~yr, $E = 10^{51}\;\erg$, 
$M = 10 \Msol$). In each of the panels on the left the solid line shows the
mass fraction of evaporated material, the dashed line shows the ejecta mass
fraction, and the dotted line shows the swept-up mass fraction. In each of the
panels on the right the dashed line shows the kinetic energy fraction, 
and the dotted line shows the thermal energy fraction, both in terms of
the {\em current} remnant energy. The solid line shows the total energy 
as a fraction of the {\em initial} remnant energy. Note that we subtract 
the thermal energy swept-up by the remnant - if this is included both 
the total energy and the thermal energy fraction increase at late times.
The vertical lines mark the ends of the free expansion (FE) and 
Quasi-Sedov-Taylor (QST) phases.}
\label{fig:fig3}
\end{figure*}

\begin{figure*}
\begin{center}
\psfig{figure=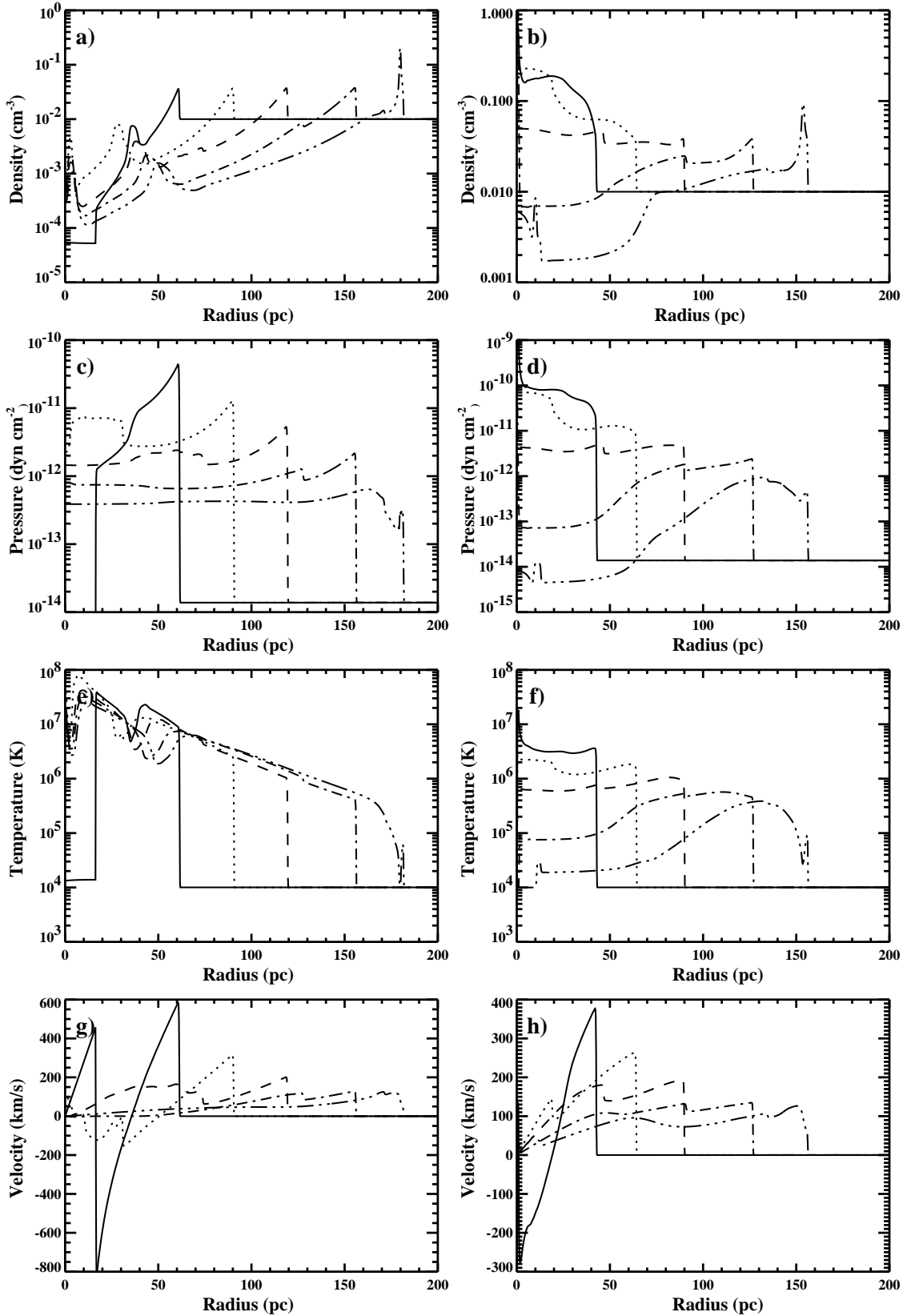,width=15.0cm}
\end{center}
\caption[]{Remnant evolution of density, pressure, temperature, and velocity
as a function of age for $n_{0}=0.01$ and $f'=0$ (left panels) or $f'=3.16$
(right panels). The profiles are at time t=$3.54 \times 10^{4}\;{\rm yr}$ 
(solid); t=$8.59 \times 10^{4}\;{\rm yr}$ (dotted); 
t=$1.70 \times 10^{5}\;{\rm yr}$ (dashed);
t=$3.39 \times 10^{5}\;{\rm yr}$ (dot-dashed); 
t=$5.41 \times 10^{5}\;{\rm yr}$ (dot-dot-dot-dashed). 
($t_{\rm sf} = 4 \times 10^{5}$~yr).}
\label{fig:fig4}
\end{figure*}

\begin{figure*}
\begin{center}
\psfig{figure=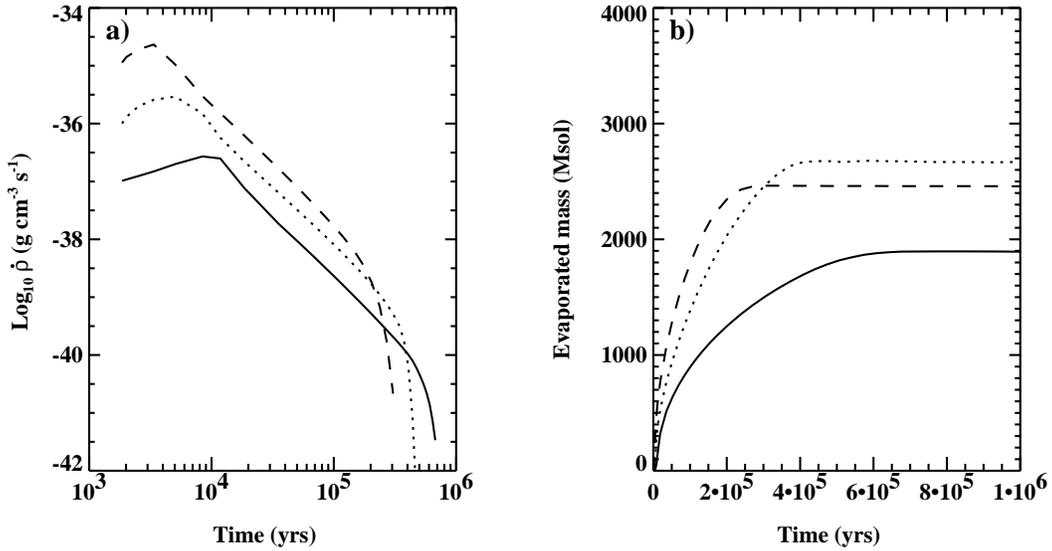,width=15.0cm}
\end{center}
\caption[]{Remnant averaged mass injection rate {\bf a)}, 
and integrated mass {\bf b)}, as a function of time for 
$n_{0} = 0.01\;\pcm3$ and $f'=3.16$ (solid), $f'=31.6$ (dotted), and $f'=316$
(dashed). Here 
we define the remnant volume as a sphere with radius equal to that of the 
blast wave. The slight increase in $\dot{\rho}$ at early time results from the fact that
the post-shock region where mass loading occurs is above the 
saturation temperature, while the volume filling factor simultaneously 
increases as the reverse shock propagates back to the centre of the remnant.
Note also that the maximum amount of mass that can be evaporated when
$n_{0} = 0.01\;\pcm3$ is $\ltsimm 3000\;\Msol$.}
\label{fig:fig5}
\end{figure*}

\begin{figure*}
\begin{center}
\psfig{figure=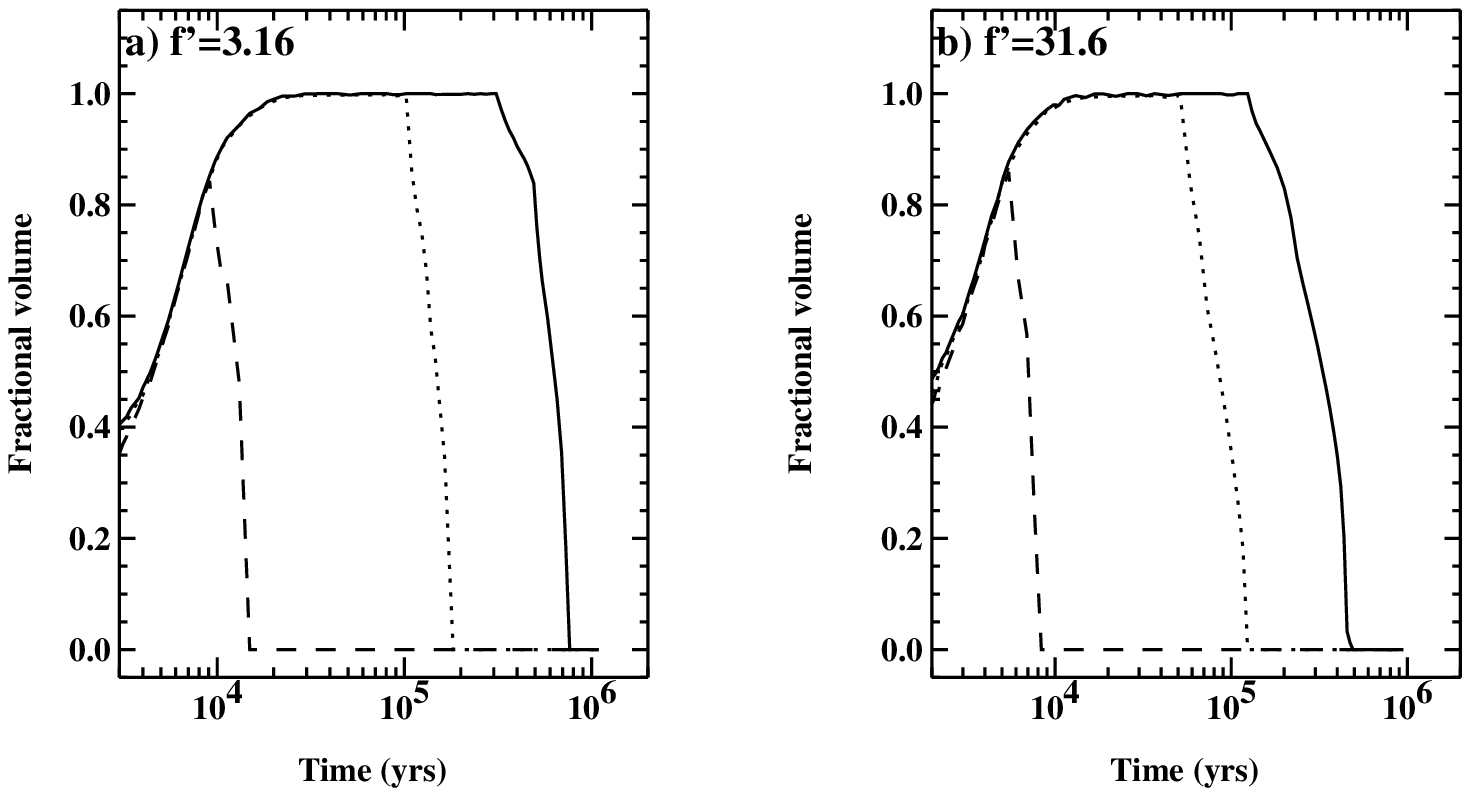,width=15.0cm}
\end{center}
\caption[]{The fractional volume of the remnant above a temperature of
$10^{5}$~K (solid), $10^{6}$~K (dotted), and $10^{7}$~K (dashed) for
$n_{0} = 0.01\;\pcm3$ and $f'=3.16$ {\bf a)} or $f'=31.6$ {\bf b)}.
At early times when the ejecta is being thermalized all of the 
shocked gas is hotter than $10^{7}$~K and the curves are coincident in
each case, and rising towards a value of 1.0. When the mass loading
is stronger the remnant cools quicker.}
\label{fig:fig5.5}
\end{figure*}

\section{Discussion}
\label{sec:discussion}

\subsection{The Free Expansion (FE) Phase}
\label{sec:discuss_fe}

Table~\ref{tab:mass_fe} gives the total mass in the remnant at the end of the
FE phase, $M_{\rm FE}$, for remnants with an initial energy, 
$E = 10^{51}\;\erg$, and ejecta mass, $M = 10 \;\Msol$. 
When there is no mass loading the total mass is
independent of the ambient density (being approximately $56 \;\Msol$, \cf 
Paper~I), though with intermediate rates of mass loading 
(\eg $f' \approx 0.316$) $M_{\rm FE}$ becomes 
dependent on $n_{0}$. For slightly higher mass loading rates ($f' \geq 3.16$)
$M_{\rm FE} \approx 37 \;\Msol$, and is insensitive to both $f'$ and 
$n_{0}$. The fact that $M_{\rm FE}$ is dependent on $f'$ for a given $n_{0}$
is consistent with Figs.~\ref{fig:fig3} 
(b,d,f), where it can be seen that the remnant energy thermalizes more
rapidly with increasing $f'$. This behaviour  
arises from the fact that mass loading reduces the rate at which
the remnant expands at early times, as it causes more braking in the ejecta and
sends the reverse shock back through it more quickly. In contrast, we 
note that $M_{\rm FE}$ is essentially constant when mass loading occurs 
by hydrodynamic ablation (Paper~I). Since the mass loading is essentially 
saturated in both formulations at this stage,
these differences arise from the fact that in this
paper we also mass load in the shocked ejecta (unlike in Paper~I), which has 
the effect of increasing its pressure.

In Fig.~\ref{fig:fig7} we show the radius at the end of the
free-expansion stage, $R_{\rm FE}$, as a function of $n_{0}$ and $f'$.
We derive an appropriate analytical approximation for
the dependence of $R_{\rm FE}$ on $n_{0}$ and $f'$ 
by following the same procedure as in Sec.~4.1 of Paper~I.  
The agreement between
the simulations and the analytical approximation is generally good, 
although the latter systematically underestimates $R_{\rm FE}$ at high 
values of $f'$ (as was also found in Paper~I, where an explanation for 
this is given). 
For future purposes, however, we are more concerned with accurate estimates
for $R_{\rm QST}$, the radius at the end of the Quasi-Sedov-Taylor phase.
We note that the much higher values of $f$ which we consider in this paper
lead to a greater range in $R_{\rm FE}$ at a given $n_{0}$.

\subsection{The Quasi-Sedov-Taylor (QST) Phase}
\label{sec:discuss_qst}

In Fig.~\ref{fig:fig8} and Table~\ref{tab:rad_qst} we show the variation 
of $R_{\rm QST}$ with $n_{0}$ and $f'$. We 
find that the expression

\begin{equation}
\label{eq:r_qst}
R_{\rm QST} (f',n_{0}) \approx 0.75 R_{\rm QST} (f'=0) f'^{-0.05} 
10^{-b\;{\rm tanh}(n_{0}/a)},
\end{equation}

\noindent where $a = 6 \times 10^{-0.9(1-e^{-x})x}$, 
$b = 0.17\;({\rm log_{10}}f' + 2.5)^{1.1}$, and 
$x= {\rm log_{10}}f' + 0.5$ achieves a good
fit to the numerical results. We plot values of $R_{\rm QST}$ using
Eq.~\ref{eq:r_qst} in Fig.~\ref{fig:fig8}.

\begin{figure}
\begin{center}
\psfig{figure=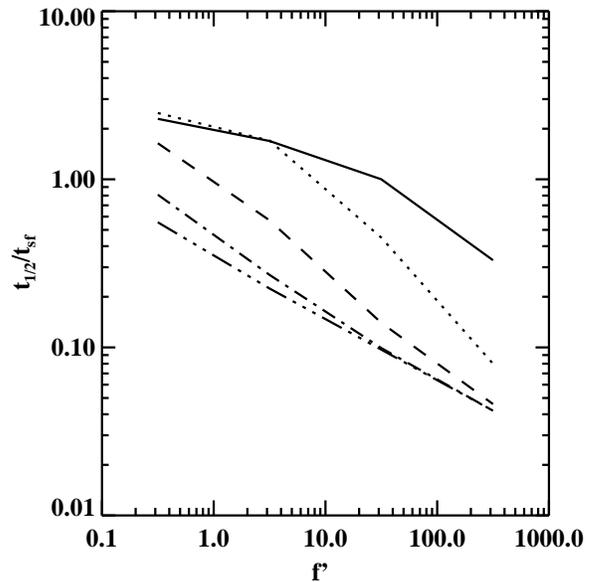,width=8.8cm}
\end{center}
\caption[]{Ratio of the half energy time to the thin shell formation time
($t_{1/2}/t_{\rm sf}$) as a function of the mass loading parameter $f'$ for 
the full range of intercloud ambient densities considered: $n_{0} = 0.01$ 
(solid); $n_{0} = 0.1$ (dotted); $n_{0} = 1.0$ (dashed); $n_{0} = 10.0$ 
(dot-dashed); $n_{0} = 100.0$ (triple dot-dash).}
\label{fig:fig6}
\end{figure}

\begin{figure}
\begin{center}
\psfig{figure=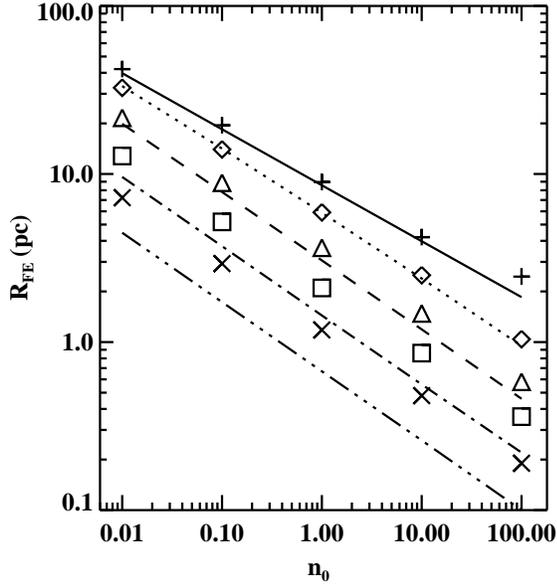,width=8.8cm}
\end{center}
\caption[]{Radius at the end of the free expansion stage as a function of
ambient density. The symbols represent the results of the numerical 
simulations and the lines are from Eqs.~8 and~9 in Paper~I, appropriately modified
for $M_{\rm FE,avg} \approx 47 \;\Msol$, and using the relation 
$f' = 316 f$. The symbols and lines are: 
$f' = 0.0$ (plus, solid); $f' = 0.316$ (diamond, dotted); $f' = 3.16$ 
(triangle, dashed); $f' = 31.6$ (square, dot-dashed); $f' = 316$ (cross, 
triple dot-dashed).}
\label{fig:fig7}
\end{figure}

\begin{figure}
\begin{center}
\psfig{figure=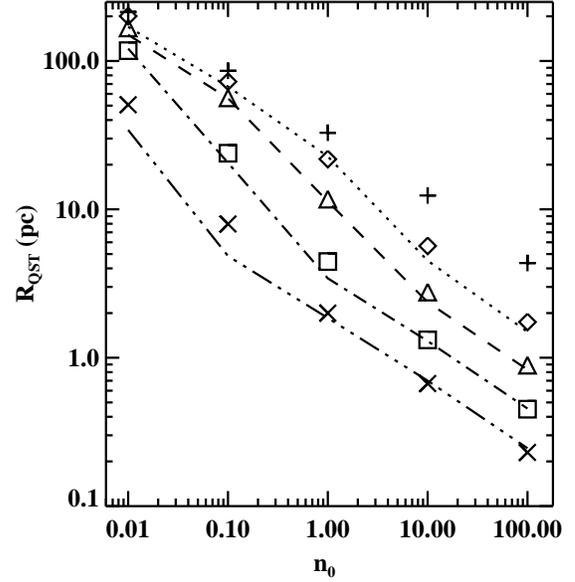,width=8.8cm}
\end{center}
\caption[]{Radius at the end of the Quasi-Sedov-Taylor stage as a function of
$n_{0}$ and $f'$. The symbols represent the results of the numerical 
simulations and the lines are the analytic approximation of Eq.~\ref{eq:r_qst},
and denote $f' = 0.0$ (plus); $f' = 0.316$ (diamond, dotted); 
$f' = 3.16$ (triangle, dashed); $f' = 31.6$ (square, dot-dashed); 
$f' = 316$ (cross, triple dot-dashed).}
\label{fig:fig8}
\end{figure}

\begin{figure}
\begin{center}
\psfig{figure=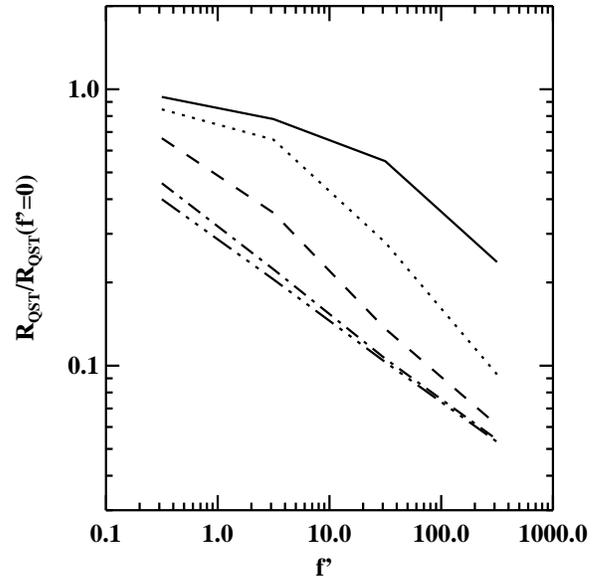,width=8.8cm}
\end{center}
\caption[]{Ratio of the radius at the end of the Quasi-Sedov-Taylor phase
to the radius at the same stage for the case where $f'=0$ 
($R_{\rm QST}/R_{\rm QST}(f'=0)$) against $f'$ for all $n_{0}$: $n_{0} = 0.01$ 
(solid); $n_{0} = 0.1$ (dotted); $n_{0} = 1.0$ (dashed); $n_{0} = 10.0$ 
(dot-dashed); $n_{0} = 100.0$ (triple dot-dashed).}
\label{fig:fig9}
\end{figure}

The ratio of the radius at the end of the Quasi-Sedov-Taylor phase
to the radius at the same stage for the case where $f'=0$ 
($R_{\rm QST}/R_{\rm QST}(f'=0)$) is shown in Fig.~\ref{fig:fig9} 
for all $f'$ and $n_{0}$. Although there is variation with $n_{0}$ for
$n_{0} \leq 10$, this ratio appears to become fairly insensitive to 
$n_{0}$ for larger values of $f'$. We further note that the behaviour seen in
Figs.~\ref{fig:fig6} and~\ref{fig:fig9} is qualitatively similar (which is
also the case for ablative mass loading \cf Paper~I). 

For an ambient density, $n_{0} = 0.01$, the approximation noted in Eq.~11
of Paper~I, with substitution of $f$ with $f'$, \ie

\begin{equation}
\label{eq:r_qst_v2}
R_{\rm QST} \approx 0.9 R_{\rm QST}(f'=0) (10\sqrt{f'})^{-0.09{\rm log_{10}}f'},
\end{equation}

\noindent is a good fit, while for high ambient densities ($n_{0} \sim 100.0$)
an excellent approximation is

\begin{equation}
\label{eq:r_qst_v3}
{\rm log_{10}} R_{\rm QST} \approx -0.3 {\rm log_{10}}f' - 0.5.
\end{equation}

\noindent In the equations in this section it is again assumed that the 
remnants have an initial energy, $E = 10^{51}\;\erg$, and 
ejecta mass, $M = 10 \;\Msol$.

\section{Conclusions}
\label{sec:conclusions}

We have extended work presented in Paper~I on the evolution of mass loaded
supernova remnants by considering mass loading by conductively driven 
evaporation of
embedded clouds. Paper~I confirmed that mass injection can strongly influence
remnant evolution, and we also establish here that the {\em nature}
of the mass injection process is also important in this regard. This 
is due to the fact that mass loading through conductive evaporation is
extinguished at relatively early times. Hence, conductively driven mass loading
does not appreciably alter the later stages of remnant evolution, when the
remnant is dominated by swept up ambient gas. Therefore remnants that ablatively mass 
load are dominated by loaded mass and thermal energy at late times (Paper I), while 
those that conductively mass load are dominated by swept-up mass and 
kinetic energy. The greater dominance of loaded mass in the ablative case
means that such remnants evolve more quickly, and reach all dynamical 
stages earlier. At a given age they tend to be both more massive and smaller
than equivalent remnants which are conductively mass loaded. 

We are able to confirm some of the properties of conductively mass loaded
remnants predicted from self-similar solutions, and in particular find that 
such remnants may display a thick-shell morphology (\cf the hydrodynamic
results presented in Cowie \etal \cite{CMO1981} and the similarity solutions
presented in Dyson \& Hartquist \cite{DH1987}).

In this work we have been particularly interested in the range
of conductively mass loaded supernova remnants at the time at which
they have radiated away half of their initial energy (see Table~\ref{tab:rad_qst}).
It was noted in Sec.~\ref{sec:calcs} that $f'$ may be dependent
on $n_{0}$. This behaviour would pick-out 
a roughly diagonal line in Table~\ref{tab:rad_qst}, and implies 
that the radius of remnants at the end of the quasi-Sedov-Taylor stage 
has less variance with $n_{0}$ than would otherwise be the case. However,
since $f'$ is also dependent on the number density of cold clouds we 
do not expect a particularly tight relationship.

Simple approximations that fit the evolution of the range of supernova 
remnants which conductively mass load, and which are complementary to 
similar approximations in Paper~I, have also been found. In both works it
is assumed that the remnant has an initial energy, $E = 10^{51}\;\erg$, 
and an ejecta mass, $M = 10 \Msol$. We expect that the evolution of remnants 
which ablatively mass load will be fairly insensitive to the progenitor 
mass, because the majority of the mass loading occurs after the FE stage
ends, at which point the swept up mass is about 6 times greater than the
progenitor mass. In the conduction case it is not so clear what will happen,
because most of the mass loading occurs in the early stages
when the remnant is becoming thermalized. Furthermore, the mass loading may
well be more dependent on our model assumptions (such as the saturation 
temperature, $T_{\rm sat}$), than on the progenitor mass, at least in 
some regions of parameter space. With regards to the possibility of 
different explosion energies, we note that our solutions should scale in
a similar way to the time of thin shell formation, 
$t_{\rm sf} \propto E^{3/14}$ (\cf Eq.~\ref{eq:tsf}).

Our range approximations will form the basis of future work to 
investigate galactic superwinds formed by the combination of many 
overlapping supernovae.
 
\begin{acknowledgements}
We would like to thank Dr. R. Bandiera for constructive comments which
led to clarification of our assumptions and generally improved the paper.
JMP would also like to thank PPARC for the funding of a PDRA position. 
This work has made use of NASA's Astrophysics Data System Abstract Service.
\end{acknowledgements}


\begin{thebibliography}{}
\bibitem[1991]{AF1991} Arthur, S.~J., Falle, S.~A.~E.~G., 1991, MNRAS, 251, 93
\bibitem[1993]{AF1993} Arthur, S.~J., Falle, S.~A.~E.~G., 1993, MNRAS, 261, 681 
\bibitem[1996]{AH1996} Arthur, S.~J., Henney, W.~J., 1996, ApJ, 457, 752
\bibitem[1995]{CB1995} Chevalier, R.~A., Blondin, J.~M., 1995, ApJ, 444, 312
\bibitem[1981]{CL1981} Chi\`{e}ze, J.~P., Lazareff, B., 1981, A\&A, 95, 194
\bibitem[1999]{C1999} Chu Y.-H., \etal, 1999, IAU Symp. 190, New Views of the
Magellanic Clouds, ed. Y.-H. Chu, N. Suntzeff, J. Hesser, D Bohlender, 143  
\bibitem[1988]{CMB1988} Cioffi, D.~F., McKee, C.~F., Bertschinger, E., 1988,
ApJ, 334, 252 
\bibitem[1977]{CM1977} Cowie, L.~L., McKee, C.~F., ApJ, 211, 135
\bibitem[1981]{CMO1981} Cowie, L.~L., McKee, C.~F., Ostriker, J.~P., 
1981, ApJ, 247, 908
\bibitem[1977]{CS1977} Cowie, L.~L., Songalia, A., 1977, Nature, 266, 501
\bibitem[2002]{DAH2002} Dyson, J.~E., Arthur, S.~J., Hartquist, T.~W., 2002, A\&A, 390, 1063 (Paper~I)
\bibitem[1987]{DH1987} Dyson, J.~E., Hartquist, T.~W., 1987, MNRAS, 228, 453
\bibitem[1982]{FG1982} Falle, S.~A.~E.~G., Garlick, A.~R., 1982, MNRAS, 201, 635
\bibitem[1996]{FK1996} Falle, S.~A.~E.~G., Komissarov, S.~S., 1996, MNRAS, 278, 586
\bibitem[1998]{FK1998} Falle, S.~A.~E.~G., Komissarov, S.~S., 1998, MNRAS, 297, 265
\bibitem[1994]{FMATT1994} Franco, J., Miller, W.~W., Arthur, S.~J., 
Tenorio-Tagle, G., Terlevich, R., 1994, ApJ, 435, 805
\bibitem[1991]{FTBR1991} Franco, J., Tenorio-Tagle, G., Bodenheimer, P., 
Rozyczka, M., 1991, PASP, 103, 803
\bibitem[1973]{G1973} Gull, S.~F., 1973, MNRAS, 161, 47
\bibitem[1986]{HDPS1986} Hartquist, T.~W., Dyson, J.~E., Pettini, M., 
Smith, L.~J., 1986, MNRAS, 221, 715
\bibitem[1977]{MC1977} McKee, C.~F., Cowie, L.~L., 1977, ApJ, 215, 213
\bibitem[1977]{MO1977} McKee, C.~F., Ostriker, J.~P., 1977, ApJ, 218, 148
\bibitem[2001]{PDFH2001} Pittard, J.~M., Dyson, J.~E., Falle, S.~A.~E.~G.,
Hartquist, T.~W., 2001, A\&A, 375, 827
\bibitem[1996]{S1996} Suchkov, A.~A., Berman, V.~G., Heckman, T.~M., 
Balsara, D.~S., 1996, ApJ, 463, 528
\bibitem[1992]{TTFM1992} Terlevich, R., Tenorio-Tagle, G.,  
Franco, J., Melnick, J., 1992, MNRAS, 255, 713
\bibitem[1999]{TM1999} Truelove, J.~K., McKee, C.~F., 1999, ApJS, 120, 299
\bibitem[2002]{WHS2002} Warren, J.~S., Hughes, J.~P., Slane, P.~O., 2003, 
ApJ, 583, 260
\bibitem[1991]{WL1991} White, R.~L., Long, K.~S., 1991, ApJ, 373, 543
\end{thebibliography}
\end{document}